\documentclass[twocolumn,a4paper,nofootinbib,showpacs,aps,floatfix,superscriptaddress]{revtex4-1}
\usepackage{color}
\usepackage{graphicx}
\usepackage{bm}
\newcommand{\ignore}[1]{}
\def\beq{\begin{equation}}
\def\eeq{\end{equation}}
\def\beqa{\begin{eqnarray}}
\def\eeqa{\end{eqnarray}}

\begin{document}
\title{Dynamic generation of spin-squeezed states in bosonic Josephson junctions}

\author{B. Juli\'a-D\'{\i}az~$^1$, 
T. Zibold~$^3$, 
M. K. Oberthaler~$^3$, 
M. Mel\'e-Messeguer~$^2$, 
J. Martorell~$^2$, 
and 
A. Polls~$^2$}
\affiliation{$^1$ ICFO-Institut de Ci\`encies Fot\`oniques, 
Parc Mediterrani de la Tecnologia, 08860 Barcelona Spain\\
%}
%\affiliation{
$^2$ Departament d'Estructura i Constituents de la Mat\`{e}ria,
Universitat de Barcelona, 08028 Barcelona, Spain\\
%}
%\affiliation{
$^3$ Kirchhoff Institute for Physics, 
University of Heidelberg, INF 227, 69120 Heidelberg, Germany}

\date{\today}
\begin{abstract}
We analyze the formation of squeezed states in a 
condensate of ultracold bosonic atoms confined by 
a double-well potential. The emphasis is set on 
the dynamical formation of such states from 
initially coherent many-body quantum states. Two 
 cases are described: the squeezing 
formation in the evolution of the system around 
the stable point, and in the short time evolution 
in the vicinity of an unstable point. The latter 
is shown to produce highly squeezed states on very 
short times. On the basis of a semiclassical 
approximation to the Bose-Hubbard Hamiltonian, we 
are able to predict the amount of squeezing, its 
scaling with $N$ and the speed of coherent spin 
formation with simple analytical formulas which 
successfully describe the numerical Bose-Hubbard 
results. This new method of producing highly 
squeezed spin states in systems of ultracold atoms 
is compared to other standard methods in the literature.   
\end{abstract}

\maketitle

\section{Introduction}
\label{sec1}

Condensates of ultracold atoms provide an exceptional 
tool to understand and control a number of phenomena 
in the fields of condensed matter, many-body quantum 
mechanics and quantum 
information/computation~\cite{lewenstein-adv,bloch-rmp}. 
Condensates are bosonic many-body quantum systems whose 
Hamiltonian can be tuned via Feshbach resonance 
techniques or by varying the trapping conditions. 

In particular we shall be interested here in condensates 
of ultracold bosonic atoms trapped in an external 
double-well potential, thus giving rise to the so-called 
external Josephson dynamics~\cite{Albiez05, GO07, stei07, esteve08, gross10}. 
The case of atoms with two internal states trapped in 
a common harmonic potential is similar. In this case 
the Josephson dynamics takes place between the two 
internal states~\cite{zib10}. A first relevant 
observation for these systems was that of the predicted 
self-trapped regime~\cite{Sme97, Mil97}, which appears 
already in the semiclassical description of the two-site 
Bose-Hubbard Hamiltonian. Later, the emphasis has been 
set on producing strongly correlated quantum states 
with appealing quantum properties such as entangled 
states~\cite{kita,korb05,esteve08}, or squeezed states 
with possible application in quantum 
metrology~\cite{wine92,wine94,gross10,spinrev}. 
Recently, the limits imposed by finite temperature 
on the maximal attainable spin squeezing have been 
discussed in Ref.~\cite{sinatra11}. 

Most of the studies have concentrated on quantum many-body 
properties present in the ground state. Notably studying 
the possibility of having cat-like many-body ground 
states~\cite{cirac98,jame05, ST08, ours10,rela} or largely 
squeezed states~\cite{esteve08}. In this paper we focus 
on the dynamical generation of squeezed states, that is, 
we consider a condensate initially prepared in a coherent 
state which is left to evolve in a suitable Hamiltonian 
so as to give rise to entangled many-body states during 
the time evolution. Our aim is thus to build those particular 
states from initial states that can be constructed with 
present experimental techniques. We will use the Bose-Hubbard 
Hamiltonian to study numerically the time evolution 
by solving the corresponding time dependent Schr\"odinger 
equation (TDSE). Alternatively, we apply a semiclassical 
approximation (based on a perturbative expansion in 
$1/N$, $N$ the number of atoms) to obtain simple and 
yet accurate expressions describing the dynamics of the 
relevant expectation values. Similar methods have been 
used in recent years to study the thermodynamic 
limit of the Lipkin-Meshkov-Glick model~\cite{lipkin}, 
which can be mapped into the usual two-site Bose-Hubbard, 
finding exact expressions for the ground state in the 
thermodynamic limit~\cite{vidal1} and characterizing 
entanglement properties of the ground state in the same 
limit~\cite{vidal2}.

The article is organized as follows, first we introduce 
the Bose-Hubbard (BH) Hamiltonian in Sect.~\ref{sec2}, 
and give a short reminder of the semiclassical approximation 
in Sect.~\ref{sec3}. In Sects.~\ref{sec4} and~\ref{sec5}, 
we propose an experimentally feasible setup for producing 
dynamically a new kind of squeezed states and study their 
properties. A comparison with the adiabatic and diabatic 
one-axis squeezing is presented in Sec.~\ref{sec7}.
In Sect.~\ref{sec6} we outline our conclusions.

\section{Two site Bose-Hubbard Hamiltonian and squeezing}
\label{sec2}

Let us consider a many-body system of bosons described 
by a two-site Bose-Hubbard Hamiltonian of the form 
$\hbar H_{\rm BH}$ with
\beq
H_{\rm BH} = -J (a_1^{\dag} a_2 + a_2^{\dag} a_1) 
+  {U\over 2}\left( \hat{n_1}(\hat{n_1}-1) 
                  + \hat{n_2}(\hat{n_2}-1)\right) \,,
\label{bh01}
\eeq
where ${\hat n}_i = a_i^{\dag} a_i$, and 
$[a_i,a_j^{\dag}] =\delta_{i,j}$. $J$ is the hopping 
strength, taken positive, and $U$ is the non-linear 
coupling strength. $U>0$ and $U<0$ correspond to 
repulsive and attractive interactions, respectively. 
To remain close to ongoing experimental realizations 
we will concentrate on the case of repulsive interactions 
among the atoms. The time dependent Schr\"odinger 
equation is written as, 
\beq
{\imath}  \partial_t |\Psi\rangle = H_{\rm BH} |\Psi\rangle \,.
\label{tdbh}
\eeq
An appropriate many-body basis for this bosonic system 
is the Fock basis~\cite{Leggett01}, 
$\{ |N_1, N_2 \rangle\}$, with $N_1+N_2=N$. Since 
the total number of atoms, $N$, is taken to be 
constant it will be more convenient to introduce 
a different notation: $N_1=k$, $N_2=N-k$. A general 
many-body state, $|\Psi\rangle$, can then be written 
in this basis as, 
\beq
|\Psi\rangle = \sum_{k=0}^N c_k |k, N-k\rangle \,.
\label{eq:ps1}
\eeq
The low energy stationary states of the system are 
characterized by values of $c_k$ that vary smoothly 
with $k$ and that take vanishingly small values when 
$k \to 0$ or to $\to N$, which corresponds to 
negligible probabilities for finding almost all the 
atoms on one of the two sites.

It is customary to define three operators 
${\bf \hat{J}}\equiv (\hat{J}_x,\hat{J}_y,\hat{J}_z)$~\cite{Leggett01, holtaus01}
\begin{eqnarray}
{\hat J}_x &=& \frac{1}{2} (a_1^{\dag} a_2 + a_2^{\dag} a_1)\nonumber \\
{\hat J}_y &=& \frac{1}{2i}(a_1^{\dag} a_2 - a_2^{\dag} a_1)\nonumber\\
{\hat J}_z &=& \frac{1}{2} (a_1^{\dag} a_1 - a_2^{\dag} a_2)\,.
\label{eq:in1}
\end{eqnarray}
In terms of these, the Hamiltonian reads 
\beq
H_{\rm BH}= -2J \hat{J}_x + U \hat{J}_z^2 
+ U\left({\hat{N}^2\over 4}-{\hat{N}\over 2}\right)\,.
\label{eq:in2}
\eeq
An important consequence of the form of this Hamiltonian 
is the existence of squeezed spin eigenstates in the 
Fock representation~\cite{kita}. This pseudo-spin 
is the one defined in Eq.~(\ref{eq:in1}). These 
states are of special importance as they incorporate 
correlations which are beyond mean-field.

Here instead we will study the dynamical generation 
of squeezing: we assume that at $t=0$ the system is 
initially prepared in a coherent state characterized by 
$(\theta,\phi)$~\cite{holtaus01}:
\beq
|\Psi_{\theta,\phi}\rangle =
 \sum_k \left(\matrix{N \cr k}\right)^{1/2} 
(\cos \theta/2)^k (e^{\imath \phi}\sin \theta/2)^{N-k} |k,N-k\rangle \,,
\nonumber 
\label{eq:cohs}
\eeq
which corresponds to a state in which all atoms populate 
the same single particle state, $\cos(\theta/2) |1\rangle
+ {\rm e}^{i \phi} \sin(\theta/2) |2\rangle $ where 
$|1\rangle = a_1^\dagger |{\rm vac}\rangle$ and 
$|2\rangle = a_2^\dagger |{\rm vac}\rangle$.  
Such states have been recently engineered, producing 
and characterizing them in a wide range of values of 
$(\theta,\phi)$~\cite{zib10}. 

Moreover, coherent states have simple expectation values 
of $\hat{J}_x$, $\hat{J}_y$, and $\hat{J}_z$~\cite{holtaus01}, 
\beqa
\langle \Psi_{\theta,\phi}|\hat{J}_x |  \Psi_{\theta,\phi}\rangle 
&=& {N\over2}
\sin{\theta}\cos{\phi} \,,\nonumber \\
\langle \Psi_{\theta,\phi}|\hat{J}_y |  \Psi_{\theta,\phi}\rangle 
&=& {N\over2}
\sin{\theta}\sin{\phi}\,,\nonumber \\
\langle \Psi_{\theta,\phi}|\hat{J}_z |  \Psi_{\theta,\phi}\rangle 
&=& {N\over2}
\cos{\theta} \,,
\eeqa
which allow to represent them on the surface of a sphere 
of radius $N/2$. They can be used to define a Husimi 
distribution of any given many-body state  
$|\Phi\rangle$,~\cite{jame05} 
\beq
\rho_{\rm H}(\theta,\phi) = |\langle  \Psi_{\theta,\phi} | \Phi \rangle|^2\,.
\label{eq:husimi}
\eeq
As an example it is useful to note that the Husimi 
distribution of a coherent state characterized by 
$(\theta',\phi')$ is given by, 
\beqa
\rho_{\rm H}(\theta,\phi)&=& 
2^{-N}\left[1+\cos(\theta)\cos(\theta')\right. \nonumber \\
&+&\left.
\cos(\phi'-\phi)\sin(\theta)\sin(\theta')\right]^N\,,
\eeqa
which has a maximum of 1 for $(\theta,\phi)=(\theta',\phi')$. 

\begin{figure}[t]                         %f01
\begin{center}
\includegraphics[width=3.5cm,angle=-90]{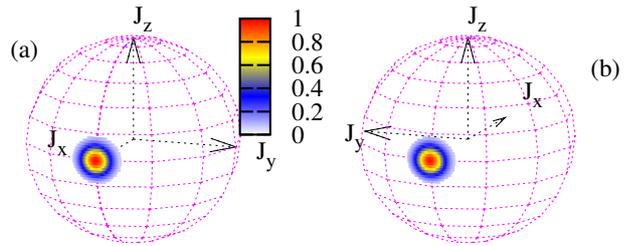}
\caption{(color online) Husimi distribution, 
$\rho_{H}(\theta,\phi)$ of the state $\Psi_{\pi/2,0}$, (a), 
and $\Psi_{\pi/2,\pi}$, (b). $N=200$.
\label{inihus}}
\end{center}
\end{figure} 
In our study we will consider as initial states two 
different coherent states: 
\beqa
|\Psi_{\pi/2,0}\rangle &=& {\cal N}_0\left(\hat{a}^\dagger_1 
+ \hat{a}^\dagger_2\right)^N |{\rm vac} \rangle
\,,\nonumber \\
|\Psi_{\pi/2,\pi}\rangle &=& {\cal N_\pi}\left(\hat{a}^\dagger_1 
- \hat{a}^\dagger_2\right)^N |{\rm vac} \rangle \,,
\label{eq:coh1}
\eeqa
with ${\cal N}_{0,\pi}$, normalization constants. The 
coefficients $|c_k|^2$ obey in both cases a binomial 
distribution 
\beq
%|c_k|^2={1\over 2^N} \left(\matrix{N \cr k}\right)\,,
|c_k|^2={1\over 2^N} {N \choose k}\,,
\label{eq:bino}
\eeq
and their Husimi distributions are, 
\beq
\rho_{\rm H}(\theta,\phi)=\left({1 \pm \cos(\phi)\sin(\theta) \over 2}\right)^N
\eeq
where the $+$ and $-$ sign corresponds to the 
$(\pi/2,0)$ and $(\pi/2,\pi)$, respectively. For large $N$, the 
equiprobability lines correspond to circles around 
$(\theta,\phi)=(\pi/2,0)$ and $(\pi/2,\pi)$, 
respectively.  The distributions are presented in 
Fig.~\ref{inihus}. 

Both initial states are especially interesting 
for two reasons: 1) they correspond to two relevant 
limiting cases which can be prepared in the laboratory. 
And 2) they 
give rise to different dynamical evolutions for $\Lambda\neq 0$. 
Starting from the $(\pi/2,0)$ state, the system  
evolves in the vicinity of a stable point in the semiclassical 
limit, producing in a natural way periodic dynamics. In 
contrast, a system  initially prepared in the 
$(\pi/2,\pi)$ state evolves in the vicinity of 
an unstable point~\cite{vardi}, in the semiclassical picture. 
That difference causes the very different maximal coherent 
squeezing found in the two cases. This will be discussed 
in greater detail in the next sections.

\subsection{Squeezing parameters}

As customary,~\cite{esteve08}, the number squeezing 
parameter is defined as, 
\beq
\xi_N^2(t)={\Delta \hat{J}_z^2 \over (\Delta \hat{J}_z^2 )_{\rm bin}}, 
\label{eq:sq}
\eeq
where 
$\Delta \hat{J}_z^2 \equiv \langle {\hat J}_z^2\rangle 
- \langle {\hat J}_z\rangle^2$ and 
$(\Delta \hat{J}_z^2 )_{\rm bin}=N/4$ in the binomial 
case~(\ref{eq:bino}). The many-body state is said to be 
squeezed if $\xi_N<1$. A second parameter which takes 
into account the coherence of the state is the so-called 
coherent spin-squeezing parameter defined 
as~\cite{wine92}~\footnote{In the cases we will consider, 
during the time evolution the wave packet remains at 
$\langle{\hat J}_{z,y}\rangle = 0$ at all times.}
\beq
\xi_S^2=  {2 J (\Delta \hat{J}_z^2)\over 
\langle \hat{J}_x\rangle^2} =  
{\xi_N^2 \over \alpha^2},
\label{eq:ssp}
\eeq
where the phase coherence is given by
\begin{eqnarray}
\alpha(t) &=& \langle \Psi(t)|{\hat \alpha}|\Psi(t) \rangle\,, \quad
{\hat \alpha} = 2 {\hat{J}_x\over N} \,.
\label{eq:co}
\end{eqnarray}
The two initial states we are considering have 
$\alpha(0)=1$ and $-1$ corresponding to $(\pi/2,0)$ 
and $(\pi/2,\pi)$, respectively. 

If a state exhibits $\xi_S<1$ it can be employed in a Ramsey type atom interferometer with an increased phase precision compared to the coherent spin state. This gain in precision can be directly related to entanglement in the system \cite{Sorensen2001}.

\subsection{Angle of maximal squeezing}

Number squeezing of a many-body state can occur along 
an axis different from the $z$ axis considered above. 
In that case one can generalize the squeezing parameter 
for an arbitrary direction 
${\bf u}\equiv (u_x,u_y,u_z)$ (${\bf u}^2=1$), as
\begin{equation}
\xi^2_{N;\bf u} = \frac{\Delta ({\bf u}\cdot {\bf \hat{J}})^2}{N/4}\,,
\label{eq:ss1}
\end{equation}
where the denominator is again the fluctuation of the 
binomial distribution. The squeezing along any direction 
in the $(y,z)$ plane only requires to calculate
$\langle ({\bf u}\cdot {\bf J})^2\rangle$ as 
$\langle ({\bf u}\cdot {\bf J})\rangle=0$. The 
corresponding generalization of the coherent spin 
squeezing parameter of Eq.~(\ref{eq:ssp}) reads, 
\beq
\xi_{S; \hat{u}}^2= {\xi^2_{N; \bf u} \over \alpha^2}\,.
\label{eq:gssp}
\eeq

As the wave packet evolves in time, there is a certain 
direction, $z'$, in which the spin squeezing is maximal. 
In a frame rotated an angle $\beta$ around the $x$ axis 
we have, 
\beqa
{\hat J}_{y'}&=& \cos \beta  {\hat J}_y + \sin \beta {\hat J}_z  \nonumber \\
{\hat J}_{z'}&=& -\sin \beta  {\hat J}_y + \cos \beta {\hat J}_z \,.
\eeqa
And since
\beqa
\langle \hat{J}_{z'}^2\rangle = 
\sin^2\beta \langle \hat{J}_y^2 \rangle 
+ 
\cos^2\beta \langle \hat{J}_z^2 \rangle
-
\sin\beta \cos\beta \langle \{\hat{J}_y,\hat{J}_z\} \rangle \,, \nonumber \\
\eeqa
requiring that $d\langle {\hat J}_{z'}^2\rangle/d\beta =0$ 
gives the angle of maximal squeezing:
\beq
\tan 2\beta_M =  
{\langle \{\hat{J}_y,\hat{J}_z\} \rangle 
\over   \langle \hat{J}_y^2 \rangle -\langle \hat{J}_z^2 \rangle}\,.
\label{bestangle}
\eeq
We will use the notation $\xi_{S;\beta_M}^2$ and 
$\xi_{N;\beta_M}^2$ for the maximal coherent spin 
squeezing and number squeezing. 

It is worth noting the role played by 
$\langle \{\hat{J}_y,\hat{J}_z\} \rangle$. If this term is 
zero, the maximal squeezing is always found either along 
$J_y$ or $J_z$. A non-zero value 
of $\langle \{\hat{J}_y,\hat{J}_z\} \rangle$ implies that 
the best squeezing will be found along a some other axis.

Eq.~(\ref{bestangle}) will allow us to compute at 
any time during the evolution the direction along 
which the squeezing is maximal. This will be of 
especial relevance for the case where the initial 
state is $|\Psi_{\pi/2,\pi}\rangle$. As will be shown 
in Section~\ref{sec5}, in this case the maximal squeezing 
gets quite sizable in the short time evolution of 
the system. 

Using the Bose-Hubbard Hamiltonian, in Sections~\ref{sec4} 
and~\ref{sec5} we will compute $\langle {\hat J}_i^2 \rangle(t)$ 
and the associated squeezing parameters for varying ratios of 
the tunneling v.s. atom-atom interaction strength, and 
present evidence for spin squeezing during the time evolution 
of the system. To better interpret these numerical results, 
we will first develop approximate expressions using a 
semiclassical model.

\section{$1/N$ approximation to the Bose-Hubbard model}
\label{sec3}

The appearance of spin squeezing in the evolution of 
the system can be studied numerically by solving the 
TDSE, Eq.~(\ref{tdbh}). It is however desirable to find 
suitable approximations which can expose the physics 
underneath the process of spin squeezing. In this 
Section we develop such approximate model and show 
that the time evolution of the system can be successfully 
mapped into the physics of a single fictitious particle 
evolving on a confining or non-confining parabolic 
potential for the $(\pi/2,0)$ or $(\pi/2,\pi)$ states, 
respectively. 
 
Following~\cite{ST08}, we introduce first an auxiliary 
Hamiltonian defined as:
\begin{eqnarray}
H_S &=& - \frac{2}{N} {\hat J}_x + \frac{U}{NJ}{\hat
  J}_z^2
=-2 h{\hat J}_x + 2 \Lambda h^2 {\hat J}_z^2 
\label{eq:s1}
\end{eqnarray}
with $h =1/N$, and $\Lambda= NU/(2J)$. It differs from 
$H_{\rm BH}$ in Eq.~(\ref{eq:in2}) in the suppression of 
the additive constants and in a factor $N J$ which makes 
it dimensionless. In the considered regime, the expectation 
values of the two terms in Eq.~(\ref{eq:s1}) are of 
similar magnitude, so that the factors $h$ compensate 
the different $N$ dependence of the expectation values 
of the two spin operators. 

\begin{figure} [t]                         %f01
\includegraphics[width=6.5cm]{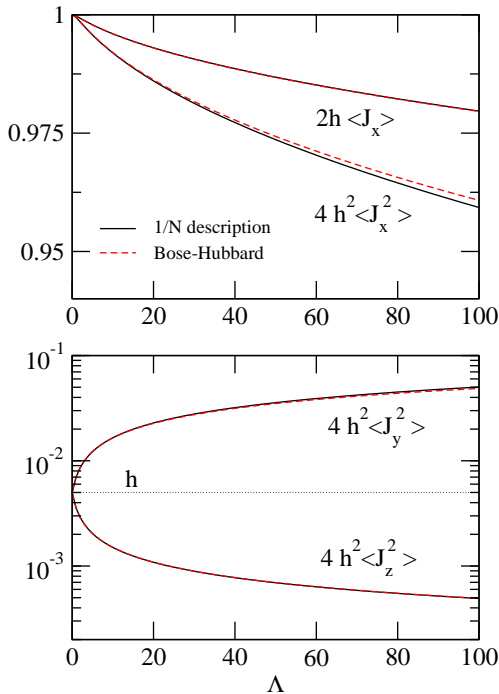}
\caption{Comparison between the ground state properties 
obtained through expressions (\ref{eq:trisapendix}), solid lines, 
and the Bose-Hubbard computation for $N=200$, 
dashed lines, as function of $\Lambda=NU/(2J)$.}
\label{fpsi1}
\end{figure} 

In Refs.~\cite{ST08, ours10-2} a semiclassical 
approximation to the TDSE has been derived. It 
uses a systematic expansion in $1/N$. Here we 
will build on this method and extend it to the 
expectation values of the quantities required 
to compute the spin squeezing and the coherence, 
Eqs.~(\ref{eq:sq},\ref{eq:co}). Earlier versions 
of the same expansion can be found also in 
Refs.~\cite{Jav99,jame05}. As explained in detail 
in Appendix~\ref{apd}, the expectation values of 
$J_x$ and $J_z^2$ can be computed from the continuous 
extension of the $c_k$'s. To deal with states close 
to the $|\Psi(\pi/2,0)\rangle$ state, Eq.~(\ref{eq:coh1}), 
one assumes that the states of interest are such 
that their $c_k$, vary smoothly: $c_k \sim c_{k\pm 1}$ 
and that the number of atoms is always large, 
$h = 1/N <<1$. This allows to introduce a 
continuous variable , $x$, and a continuous function, 
$\psi(x)$ such that 
$\psi(x=k/N) = \sqrt{N}\, c_k$~\cite{Jav99,ST08,martam,ours10-2}. 
Next a new variable $z\equiv 2 x -1$ is defined, 
and $\psi(z)$ ($-1\leq z\leq 1$), renormalized to 
$\int_{-1}^1 dz |\psi(z)|^2=1$. The expressions for 
the expectation values are,  
\beqa
h  \langle \psi| \hat{J}_x |\psi \rangle &\simeq& 
\int_{-1}^1 dz\; \psi^*(z)\,
\left[\left(\frac{h^2 \left(-1-z^2\right)}{4 \left(1-z^2\right)^{3/2}}
\right.\right.\nonumber \\
&&\left.+\frac{h}{2
  \sqrt{1-z^2}}+\frac{\sqrt{1-z^2}}{2}\right) \psi(z) 
\nonumber \\
&&\left.-\frac{h^2 z}{\sqrt{1-z^2}}\psi'(z)
+h^2 \sqrt{1-z^2} \psi''(z)\right]\nonumber \\
h^2 \langle \psi| \hat{J}_z^2 |\psi \rangle 
&=&
\int_{-1}^1 dz\; |\psi(z)|^2 \,{z^2\over 4}\,. 
\label{eq:tristext}
\eeqa
As in many other semiclassical expansions, the power 
series in $h$ is asymptotic, and one can see above 
that depending on the behavior of the chosen $\psi(z)$ 
as $z \to \pm 1$, divergent contributions will appear 
already at order $h^2$. As usual for asymptotic series 
the strategy that we will follow is to truncate those 
terms that degrade the convergence. We will detail 
later how this is done. The validity of this $1/N$ 
expansion can be seen in Fig.~\ref{fpsi1} where we 
show a comparison between our expressions and the 
exact Bose-Hubbard calculation of the ground state 
properties of the system.  

Let us now go back to the Hamiltonian, $H_S$ in 
Eq.~(\ref{eq:s1}). Using the above results, its 
semiclassical expectation value is 
\begin{eqnarray}
&&\langle \psi|H_S|\psi\rangle = 
- 2 h \langle \psi| \hat{J}_x |\psi\rangle 
+ 2 \Lambda h^2 \langle \psi | \hat{J}_z^2| \psi \rangle \,,
\label{eq:tr1}
\end{eqnarray}
When we look for the stationary points of 
$\langle \psi|H_S|\psi \rangle - E^{(s)} \langle \psi|\psi\rangle$
we arrive at, 
\begin{eqnarray}
{\cal H}_N(z) \psi(z)  &\equiv& 
-2h^2\left(\sqrt{1-z^2} \psi'' 
-\frac{z}{\sqrt{1-z^2}} \psi' \right) \label{eq:tr2} \\
&+& \left(\frac{1}{2} \Lambda z^2 - \sqrt{1-z^2} 
+ \delta {\cal V}\right)\psi(z) \nonumber \\
&\equiv&-2 h^2 \partial_z \sqrt{1-z^2}\partial_z \psi + {\cal V}(z) \psi
=  E^{(s)} \psi(z) \,,\nonumber
\end{eqnarray}
which is a pseudo-Schr\"odinger equation similar 
to the one reported in Ref.~\cite{ST08} except 
for the additional term, $\delta {\cal V}$:
\begin{eqnarray}
\delta {\cal V} &=& 
-\frac{h}{\sqrt{1-z^2} } 
+h^2 \frac{(1+z^2)}{(1-z^2)^{3/2}} 
\label{eq:tr3}
\end{eqnarray}
which was neglected in Ref.~\cite{ST08}.

Eq.~(\ref{eq:tr2}) can be regarded as a 
Schr\"odinger-like equation defined on a compact 
interval, $z\in [-1,1]$. It is expected to provide 
accurate results provided $\psi(z)$ vanishes at 
the boundaries. The equation provides an important 
insight into the problem, essentially builds on 
the semi-classical Hamiltonian, which is equal to 
${\cal V}(z)$, and quantizes it, through the effective 
mass form, $ -2h^2 \partial_z \sqrt{1-z^2}\partial_z $. 

In line with the present approximation, the time 
evolution will then be described via
\begin{equation}
\imath h \partial_{t} \psi(z,t) = {\cal H}_N \psi(z,t)\,,
\label{eq:stdse}
\end{equation}
where now $t$ is the time measured in units of $1/J$. 
The so-called ``Rabi''time of the system is 
$t_{\rm Rabi}=\pi/J$.

\section{Dynamical squeezing around a fixed stable point: $\Psi_{\pi/2,0}$ state}
\label{sec4}

\begin{figure}[t]                         %f01
\begin{center}
\includegraphics*[width=3.5cm,angle=-90]{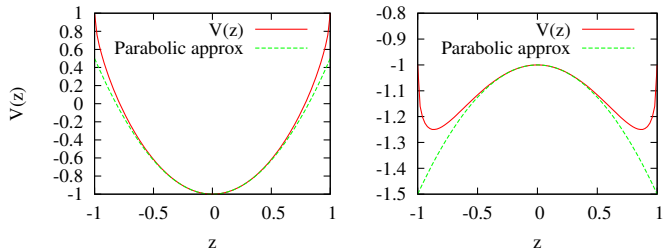}
\caption{(color online) Depiction of ${\cal V}(z)$ and 
its parabolic approximation used in Sec.~\ref{sec4} (left) and 
Sec.~\ref{sec5} (right).
\label{pot}}
\end{center}
\end{figure}

We consider now the dynamical situation where the 
condensate is initially prepared in the coherent 
state $\Psi(\pi/2,0)$, and study the squeezing 
and coherence of the system as a function of time 
as it evolves under the action of ${\cal H}_N$. 

In the limit of large $N$, small $h$, the binomial 
distribution $|c_k|^2$ corresponding to the state, 
$\Psi_{\pi/2,0}$, see Eq.~(\ref{eq:bino}), approaches 
the Gaussian distribution, in the continuous $z$ 
variable 
\beq
\psi_0(z)= \left({1\over \pi b_0^2}\right)^{1/4} 
{\rm e}^{-z^2/(2 b_0^2)}
\,,
\label{eq:gau0}
\eeq
with $b_0^2=2h= 2/N$. During the time evolution, 
$|\psi(z,t)|^2$ will be confined to a fairly narrow 
region in $z$ of size $\simeq \sqrt{2 h}$. For this 
range of values of $z$ we will approximate 
$\sqrt{1-z^2} \simeq 1$ in the kinetic energy term 
of ${\cal H}_N$, and make a parabolic approximation 
to ${\cal V}(z)$:
\beq
{\cal V}(z) \simeq  -1 -h  + {1\over 2}\, {1\over 4}\,\omega^2 z^2 
\,,
\label{eq:vapo}
\eeq
with effective mass equal to $1/4$ and frequency given 
by $\omega=2 \sqrt{1+\Lambda-h}$. Thus, the evolution of 
the $\Psi_{\pi/2,0}$ state is mapped into the evolution 
of a centered Gaussian wave packet inside a confining 
harmonic oscillator potential. The system will oscillate 
around the classical stable point, periodically building 
a certain amount of coherent spin squeezing that we 
will quantify in the following. The parabolic approximation 
is extremely accurate for our purposes. This is because 
the initial extent of the packet, 
$\sqrt{\langle z^2(0)\rangle}=\sqrt{2h}\ll 1$, is always 
the maximum value attainable during the time evolution. 

\begin{figure}[t]                         %f01
\begin{center}
\includegraphics*[width=7.5cm]{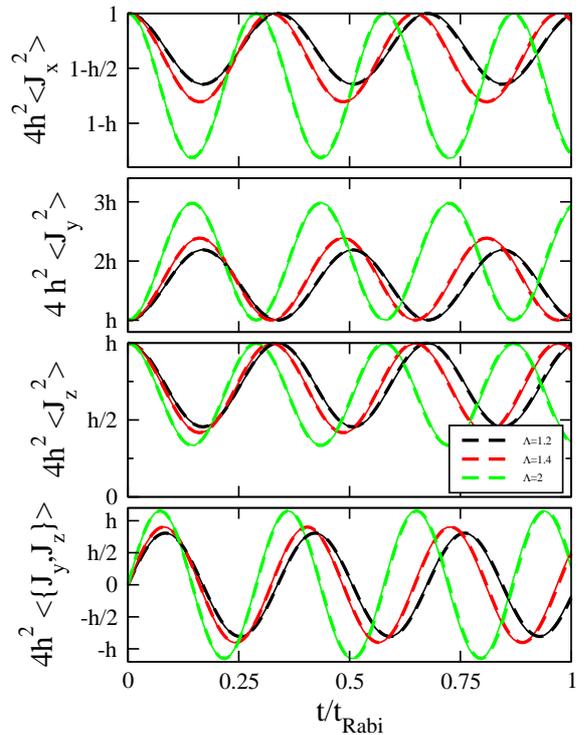}
\caption{(color online) Comparison between the 
Bose-Hubbard results, dashed lines, and the  
expressions in Eq.~(\ref{eq:trisa2}), solid lines, 
for different values of $\Lambda=1.2, 1.4$ and 2. 
The number of particles is $N=200=1/h$. The initial 
state is $\Psi_{\pi/2,0}$.
\label{tat}}
\end{center}
\end{figure} 
Under the parabolic approximation for the potential, 
see Fig.~\ref{pot}, the initial Gaussian wave packet, 
Eq.~(\ref{eq:gau0}), remains Gaussian as it evolves 
in time. The exact wave function reads, 
\beq
\psi_\Lambda(z,t) =
{1\over (\pi b^2)^{1/4}} \;
e^{i \kappa}\;
e^{-z^2 /(2 b^2)} \; 
e^{i z^2 \phi/(2 b^2)} \,,
\label{eq:sol}
\eeq
where, 
\beqa
b^2(t)&=& 
h\, \left(1+{4\over \omega^2} 
+ \left(1-{4\over \omega^2}\right)\cos 2\omega t \right)\nonumber \\
\phi(t)&=& {\omega\over 4}
\left( {4\over\omega^2}-1\right) 
\sin 2 \omega  t \,  \\
\kappa(t) &=& {1+h\over h} t +{1\over 4}
\arctan\left({\omega\over 2 \tan \omega t}\right) -{\pi\over 8} -\frac{\pi}{4}
\left[\frac{\omega t}{\pi}\right] \,,\nonumber
\eeqa
where in the last equation $[ x ]$  means integer part of
$x$. 
Now we insert the exact, $\psi_\Lambda(z,t)$ in the 
semiclassical expressions for the expectation values 
of the spin components, Eqs.~(\ref{eq:tristext}) 
and (\ref{eq:trisapendix}), and replace the denominators 
by their approximations for small $z$, 
i.e. $1/\sqrt{1-z^2} \simeq 1+ z^2/2$ or $1$ depending 
on the size of their contribution. And finally, we retain 
terms up to linear in $h$ (note that $b^2$ is proportional to $h$), 
\beqa
2h \langle \hat{J}_x \rangle &\simeq& 
1+{h\over 4}\frac{\Lambda^2}{1+\Lambda} (\cos 2 \omega t -1)
\nonumber \\
4h^2 \langle \hat{J}^2_x \rangle &\simeq& 
1+{h\over 2}\frac{\Lambda^2}{1+\Lambda} (\cos 2 \omega t -1)
\nonumber \\
4h^2 \langle \hat{J}_y^2 \rangle &\simeq&
{h\over 2} (2+\Lambda - \Lambda \cos 2 \omega t ) \nonumber
\\
4h^2 \langle \hat{J}_z^2 \rangle &\simeq&  
{h\over2(1+\Lambda)}\left( 2+\Lambda  +\Lambda 
\cos 2 \omega t \right)\nonumber \\
4 h^2\langle \{\hat{J}_y,\hat{J}_z\} \rangle &\simeq&
h {\Lambda\over \sqrt{1+\Lambda}} \sin 2 \omega t
\,.\label{eq:trisa2}
\eeqa
Within the same approximation, the angle of maximal 
squeezing, Eq.~(\ref{bestangle}), can be written as, 
\beq
\tan 2 \beta_M \simeq {2 \sqrt{1+\Lambda}\over 2+\Lambda} {1\over \tan \omega t}\,.
\label{eq:besta}
\eeq 

\begin{figure}[t]                         %f01
\begin{center}
\includegraphics[width=10cm,angle=-90]{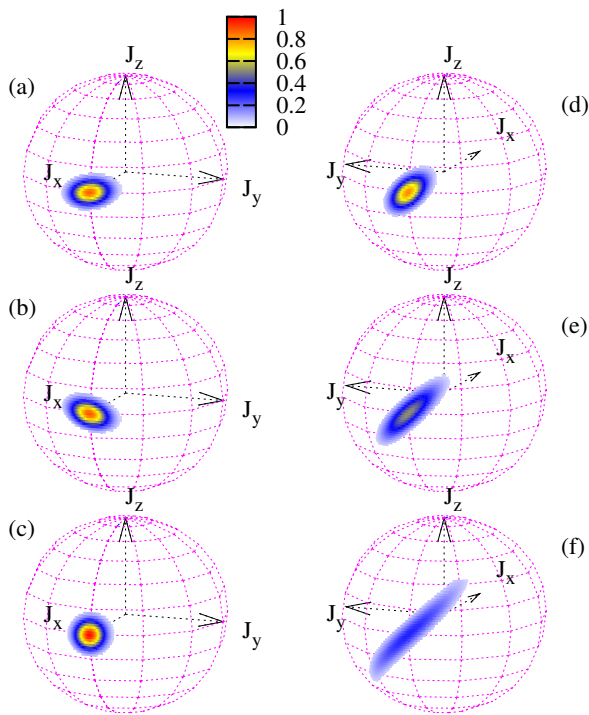}
\caption{(color online) Snapshots of the Husimi 
distribution, $\rho_{H}(\theta,\phi)$. Panels (a,b,c) and 
(d,e,f) correspond to an initial state $\Psi_{\pi/2,0}$ and 
$\Psi_{\pi/2,\pi}$, respectively. 
(a,d) are computed at $t=0.1 t_{\rm Rabi}$, (b,e) at 
$t=0.2 t_{\rm Rabi}$, and (c,f) at $t=0.3 t_{\rm Rabi}$. 
$N=200$ and $\Lambda=2$. 
\label{husirep}}
\end{center}
\end{figure} 

These approximate expressions turn out to be 
very accurate for a broad set of parameters. 
In Fig.~\ref{tat} we compare the exact Bose-Hubbard 
results and those obtained from Eqs.~(\ref{eq:trisa2}).
The initial state is $|\Psi_{\pi/2,0}\rangle$ and 
is left to evolve in a Hamiltonian with repulsive 
atom-atom interactions of $\Lambda=1.2, 1.4$ and 2. 
The expectation value of $\hat{J}_i^2$ is presented, 
$i=x,y,z$ together with the expectation value 
of $\{\hat{J}_y,\hat{J}_z\}$. As can be seen, BH 
predicts periodic oscillations for all the quantities. 
$\langle \hat{J}_x^2\rangle$ is seen to be essentially 
1 during the time evolution. The small departure from 
full coherence is well captured by the term 
$\propto h$ in the semiclassical expression. 
$\langle \hat{J}_z^2\rangle$ and $\langle \hat{J}_y^2\rangle$ 
are found to evolve in phase, as predicted 
in ~(\ref{eq:trisa2}). $\langle \{\hat{J}_y,\hat{J}_z\} \rangle$ 
is small but non-zero during the evolution, implying 
the existence of a direction along which the squeezing is maximal.

According to Eqs.~(\ref{eq:trisa2}) the wave packet 
will squeeze periodically along the $z$ direction 
with a frequency, $2 \omega$. The maximal attainable 
number squeezing takes place when $2\omega t=n \pi$, and 
is,  
\beqa
\xi_{N, {\rm max}}^2&=&{1 \over 1+\Lambda}\,.
\label{maxs}
\eeqa
Similarly we find that the coherence at maximal squeezing 
is given by, 
\beqa
\langle \hat{\alpha} \rangle_{\rm ma\; sq}&=& 
1 - h \frac{\Lambda^2}{2(1+\Lambda)}\,.
\label{comaxs}
\eeqa
The semiclassical predictions break down when the extent 
of the wave packet, $\sqrt{\langle z^2\rangle}$, is of the 
order of $h$. Using Eqs.~(\ref{eq:trisa2}) at the maximum 
number squeezing yields the condition, 
$\Lambda \lesssim 1/h=N$.

\begin{figure}[t]                         %f01
\begin{center}
\includegraphics[width=8cm,angle=0]{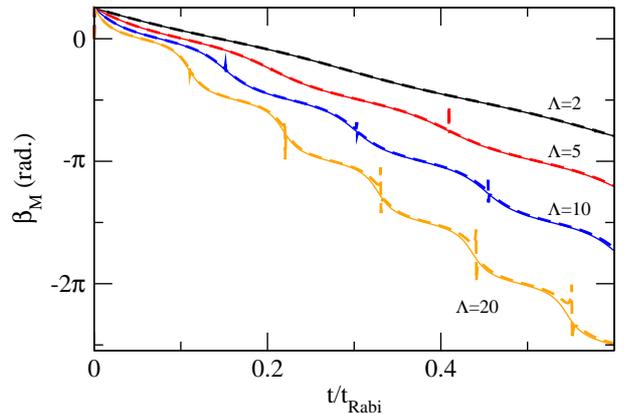}
\caption{(color online) Angle of maximal squeezing, 
Eq.~(\ref{bestangle}) computed from the Bose-Hubbard 
calculation, dashed lines, and using 
equation~(\ref{eq:besta}), solid lines. $N=400$. 
\label{angle2}}
\end{center}
\end{figure} 

As explained in the previous section a non-zero value 
of the anticommutator $\langle \{\hat{J}_y,\hat{J}_z\}\rangle$, 
as in Fig.~\ref{tat}, implies that the maximal squeezing 
is found along an axis $z'$, defined by an angle 
$\beta_M$, see Eq.~(\ref{bestangle}). This also 
reflects in the Husimi distributions depicted in 
Fig.~\ref{husirep}. In the figure we present 
three snapshots of the Husimi distributions at 
different times, $0.1, 0.2$ and $0.3$ $t_{\rm Rabi}$ 
computed for $\Lambda=2$. The Husimi distribution 
is initially symmetric, see Fig.~\ref{inihus}, 
as corresponds to a coherent state. As time evolves, 
panels (a,b,c) of Fig.~\ref{husirep}, the distribution 
is seen to be ellipsoidal but non-canonical, 
i.e. the symmetry axes of the ellipses are not 
$y$ and $z$. The angle of maximal squeezing is plotted 
in Fig.~\ref{angle2}. The angle varies almost linearly 
with time, implying that the distribution rotates 
around the $x$ direction at an almost constant velocity. 
This behavior is captured by equation (\ref{eq:besta}). 

\section{Early squeezing around an unstable point: $\Psi_{\pi/2,\pi}$ state}
\label{sec5}

\begin{figure}[t]                         %f01
\begin{center}
\includegraphics*[width=7cm,angle=-0]{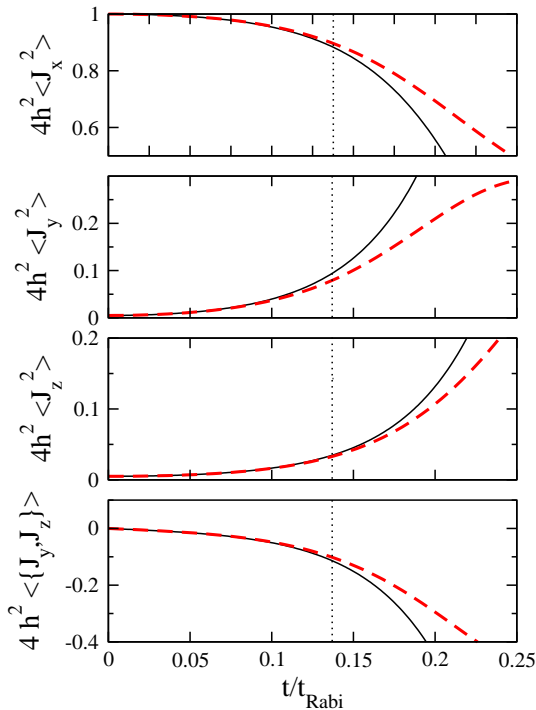}
\caption{(color online) Comparison between the exact 
Bose-Hubbard (BH) result, dashed lines, and the 
analytic expressions in Eq.~(\ref{eq:tris2b}), solid 
lines, for $\Lambda=4$. The number of particles 
is $N=200$. The initial state is $\Psi_{\pi/2,\pi}$. 
The dotted line marks the breaking of the parabolic 
approximation and is given by Eq.~(\ref{tmax}).
\label{tatt}}
\end{center}
\end{figure}

When considering the dynamics around the $(\pi/2,\pi)$ 
state in order to make use of the semiclassical model 
one has to assume that it is the $(-)^k c_k$ that vary 
smoothly. And thus introduce a continuous function 
$\psi(x=k/N) = \sqrt{N} (-1)^k c_k$~\cite{ST08,martam,ours10-2}. 
As explained in Ref.~\cite{ST08}, see also the expressions 
in our Appendix~\ref{apd}, the dynamical equation in this 
case reads, including only the lowest order in $h$ terms:
\begin{eqnarray}
\imath h \partial_t \psi(z,t) &=& \bigg( 2h^2 \partial_z \sqrt{1-z^2}
\partial_z  
\nonumber \\
&&+   \frac{1}{2}\Lambda z^2 + \sqrt{1-z^2}  \bigg)
\psi(z,t) \,,
\label{eq:sf52}
\end{eqnarray}
with a negative effective mass. For convenience we choose 
to multiply by $-1$ both sides of the equation and perform complex conjugation, so that
\begin{eqnarray}
\imath h \partial_{t} \psi^*(z,t) &=& \bigg(- 2h^2 \partial_z \sqrt{1-z^2}
\partial_z  +   {\cal V}_-(z)  \bigg) \psi^*(z,t)\,, \nonumber \\ 
\label{eq:sf52b}
\end{eqnarray}
and the evolution of $\psi^*(z,t)$ is that of an initial 
wave packet, again of the form of Eq.~(\ref{eq:gau0}), inside the 
potential, ${\cal V}_-(z) = -(1/2)\Lambda z^2  - \sqrt{1-z^2}$.
When $\Lambda >1$, this is a double-well potential in 
the $z$-space, see Fig.~\ref{pot} (right), and has 
a central barrier. Including terms of order $h$, we approximate it as 
 \beq
{\cal V}_-(z) \simeq  -1 -h - {1\over 2}\, {1\over 4}\,\bar\omega^2 z^2 
\label{eq:parm}
\eeq
where $\bar\omega=2\sqrt{\Lambda-1+h}$. Although this 
parabolic potential is non confining, we still find that 
the solution of Eq.~(\ref{eq:sf52b}) with ${\cal V}_-(z)$ 
as in Eq.~(\ref{eq:parm}) is formally identical to 
Eq.~(\ref{eq:sol}), so that (up to a phase depending 
only on $t$),
\beq
\psi_\Lambda^*(z,t)={1\over [\pi b^2]^{1/4}}e^{- {z^2\over 2 b^2(t)}} \;
e^{i\phi(t)z^2 \over 2 b^2(t)} \ .
\eeq 
\begin{figure}[t]
\begin{center}
\includegraphics[width=7cm]{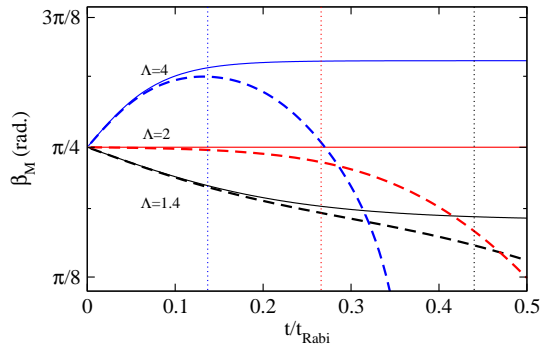}
\caption{(color online) Angle for maximal squeezing 
obtained from the Bose-Hubbard simulation, dashed lines, and 
the expression (\ref{bestangle}) with the ones in 
Eq.~(\ref{eq:tris2b}), solid lines. The dotted lines 
mark the breaking of the parabolic approximation 
for each $\Lambda$, Eq.~(\ref{tmax}). The number of particles 
is $N=200$.}
\label{angle}. 
\end{center}
\end{figure} 
\begin{figure*}[t]
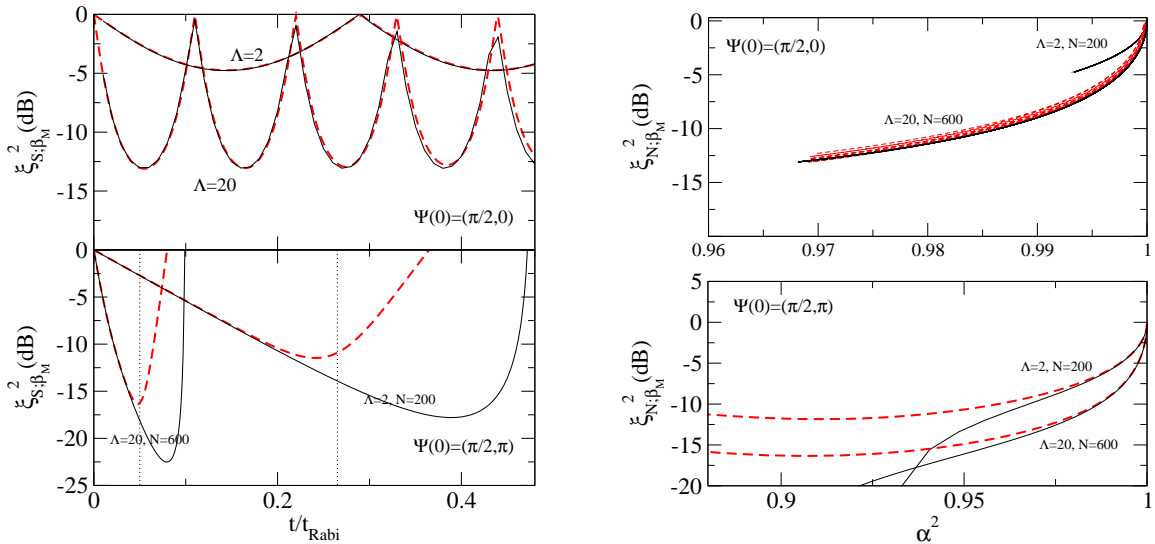
                         %f01
\begin{center}
\includegraphics[width=7cm,angle=0]{fig9a.eps}
\hspace{25pt}
\includegraphics[width=7cm,angle=0]{fig9b.eps}
\caption{(Color online) (Left) Coherent spin squeezing 
parameter, $\xi_{S;\beta_M}^2$, computed in the direction of maximal
squeezing as a function of time. The dotted lines mark the 
breaking of the parabolic approximation, Eq.~(\ref{tmax}).
(Right) Number squeezing 
parameter, $\xi_{N;\beta_M}^2$, computed in the direction of maximal
squeezing as a function of the spin coherence, $\alpha^2$. 
The upper and lower panels correspond to the initial states 
$\Psi_{\pi/2,0}$ and $\Psi_{\pi/2,\pi}$, respectively. Dashed lines 
are Bose-Hubbard calculations, while the solid 
lines are obtained using Eqs.~(\ref{eq:trisa2}) and~(\ref{eq:tris2b}). 
\label{cohsq}}
\end{center}
\end{figure*} 
However, inserting this $\psi_{\Lambda}$ in Eq.~(\ref{eq:sf52b}) 
with the parabolic approximation for ${\cal V}_-(z)$ one now 
finds:
\beqa
b^2(t)&=& 
h \left[ 1-\frac{4}{{\bar \omega}^2} + \left( 1+\frac{4}{{\bar
        \omega}^2}\right) \cosh 2 {\bar \omega} t  \right]
\nonumber \\
\phi(t)&=&  \frac{{\bar \omega}}{4} \left(\frac{4}{{\bar \omega}^2} +1\right)
        \sinh 2 {\bar \omega} t \ ,
\label{bphi2}
\eeqa
and correspondingly
\beqa
2h \langle \hat{J}_x \rangle &\simeq& 
-1+{h\over 4}\frac{\Lambda^2}{\Lambda-1} (\cosh 2 {\bar \omega} t -1)
\nonumber \\
4h^2 \langle \hat{J}^2_x \rangle &\simeq& 
1-{h\over 2}\frac{\Lambda^2}{\Lambda-1} (\cosh 2 {\bar \omega} t -1)
\nonumber \\
4h^2 \langle \hat{J}_y^2 \rangle &\simeq&
{h\over 2} (2-\Lambda + \Lambda \cosh 2 {\bar \omega} t ) 
\nonumber
\\
4h^2 \langle \hat{J}_z^2 \rangle &\simeq&  
{h\over2(\Lambda-1)}\left[ \Lambda 
(\cosh 2 {\bar \omega} t+1)-2 \right]\nonumber \\
4 h^2\langle \{\hat{J}_y,\hat{J}_z\} \rangle &\simeq&
-h {\Lambda\over \sqrt{\Lambda-1}} \sinh 2 {\bar \omega} t
\,.\label{eq:tris2b}
\eeqa
Fig.~\ref{tatt} shows that these expressions provide an 
accurate account of the short time dynamics of the 
system: Eqs.~(\ref{eq:tris2b}) predict a fast exponential 
growth of $\langle \hat{J}_{y,z}^2\rangle$, while the system 
remains mostly coherent, which agrees well with the full 
Bose-Hubbard calculation. The results suggest
that the evolution of this state will produce much 
larger squeezing, as we will quantify in the following, 
than in the case of the $(\pi/2,0)$ state, where 
$4 h^2\langle \hat{J}^2_{y,z}\rangle\sim h$. In fact, 
it will be during this short time evolution that the 
system will build its maximum coherent squeezing. Therefore 
the simple analytical predictions provide a powerful tool 
to characterize the way squeezing is produced in 
the system.

In contrast with the $(\pi/2,0)$ case, now 
 $|\psi_{\Lambda}(z,t)|^2$ gets broader in $z-$space 
during the time evolution. Thus, the simplified model 
should break down whenever the extent of the wave packet 
is comparable to the size of the allowed range for $z$: 
$\sqrt{\langle z^2\rangle}\simeq 1$, or when the 
momentum, $\hat{p}_z=-\imath h\partial_z$, is larger 
than the maximum possible, due to the underlying 
discretization, 
\beq
\sqrt{\langle \hat{p}_z^2 \rangle} \equiv 
\sqrt{\langle (-h^2\partial_{z^2})\rangle}\simeq 1/2\,.
\eeq
A good estimate of the time when the parabolic approximation 
breaks down is obtained from, 
\beq
\phi^2(t_{\rm max})\simeq 1/h,
\eeq
and thus, 
\beq
t_{\rm max} \simeq {1\over 4\bar\omega}\log\left({8 N\over \Lambda}\right)\,.
\label{tmax}
\eeq
This time predicts correctly why the parabolic 
approximation breaks down at earlier times as 
$\Lambda$ is increased.

The evolution of the many-body state is presented in 
three snapshots of its Husimi distribution in 
Fig.~\ref{husirep} (d,e,f) for $\Lambda=2$. As seen 
in Fig.~\ref{husirep} a very different behavior is 
found in comparison with the evolution of the 
$(\pi/2,0)$ state. In this case the distribution 
becomes ellipsoidal, as expected from the non-zero 
values of $\langle\{\hat{J}_y,\hat{J}_z\}\rangle$, but 
does not rotate with time.

\subsubsection*{Squeezing in the initial evolution}

As discussed above, in this case there is an 
exponential growth of $\langle \hat{J}_{y,z}^2\rangle$ 
for $t\lesssim t_{\rm max}$ . This 
feature makes this configuration very relevant for the purpose 
of producing highly squeezed states along a 
specific direction.

Inserting the semiclassical expressions given in 
Eq.~(\ref{eq:trisa2}), we get, 
\beq
\tan 2 \beta_M \simeq -2 
{\sqrt{\Lambda-1}\over \Lambda-2} \coth(\bar\omega t)
\label{tanb2}
\eeq
which for $t\lesssim t_{\rm max}$ reproduces the angle 
obtained with the full Bose-Hubbard calculation, as 
seen in Fig.~\ref{angle}. The angle at which the 
squeezing is maximal is initially $\pi/4$ regardless 
of the interaction at which the evolution is performed. 
Different values of $\Lambda$ produce evolutions in 
which either the angle grows or decreases at short 
times. From, Eq.~(\ref{tanb2}), retaining contributions 
linear in $t$ we get, 
\beq
\beta_M={\pi\over 4}-{1\over2}(2-\Lambda) t\,.
\label{eq:bana}
\eeq
Two important features seen in Fig.~\ref{angle} are 
well captured by these expressions; a) Eq.~(\ref{eq:bana}) 
predicts the angle to grow (decrease) with time for 
$\Lambda< (>) 2$, b) the value $\Lambda=2$ is predicted 
to have an almost constant angle of maximal squeezing 
for 1/4 of the Rabi time, also confirmed in the Bose-Hubbard 
calculation. 

The usefulness of the squeezing for the improvement 
of interferometric measurements is characterized by 
the two squeezing parameters introduced in Eqs.~(\ref{eq:sq}) 
and~(\ref{eq:ssp}) and their generalizations in 
Eqs.~(\ref{eq:ss1}) and~(\ref{eq:gssp}). In Fig.~\ref{cohsq} we 
depict both $\xi_{N;\beta_M}^2$ and $\xi_{S;\beta_M}^2$ computed 
along the direction of best squeezing defined in 
Eq.~(\ref{tanb2}). We compare the results obtained 
with either initial conditions considered in the article, 
$\Psi_{\pi/2,0}$, and $\Psi_{\pi/2,\pi}$. As can be seen in 
the figure, starting from the $\Psi_{\pi/2,\pi}$ the 
dynamically attainable coherent spin squeezing parameter 
is much smaller than the attainable one from the 
$\Psi_{\pi/2,0}$ state. $\xi_S^2$ remains smaller than 
one for up to $0.4$ $t_{\rm Rabi}$ for $\Lambda=2$. The 
speed of coherent spin squeezing, 
$ \partial \xi_{S;\beta_M}^2 / \partial t$ at the angle of 
best squeezing is seen to be equal when starting from 
any of the two states,  
\beq
 {\partial \xi_{S;\beta_M}^2 \over \partial t} = -2 \Lambda\,.
\label{speed}
\eeq
The maximal coherent squeezing obtained for the $(\pi/2,\pi)$ 
case is obtained at the time when the parabolic approximation 
breaks down, as seen clearly in Fig.~\ref{cohsq}. At this time 
scale, we have, 
$\tan 2 \beta_M \simeq -2 \sqrt{\Lambda-1} /( \Lambda -2)$ and 
\beq
\xi_{S,\beta_M}^2(t_{\rm max})= 2\sqrt{ {2 \over N \Lambda}} \,.
\label{maxipi}
\eeq

\section{Comparison to standard squeezing procedures}
\label{sec7}

In sections~\ref{sec4} and~\ref{sec5}, we have presented 
two methods of producing spin squeezed states. The first 
builds on the evolution of the initial state in the vicinity 
of a semiclassical stable point. The second one profits 
from the presence of a bifurcation in the semiclassical 
description. In both cases we have presented simple 
formulas which quantify how the coherent spin squeezing 
evolves with time. In this section we will compare these 
two methods to standard ones: adiabatic squeezing and 
diabatic Kitagawa-Ueda\cite{kita} one-axis twisting. 

\subsection{Adiabatic spin squeezing}

This is the maximum spin squeezing that can be 
obtained in the ground states by adiabatically varying 
the parameters of the Bose-Hubbard Hamiltonian. Experimentally 
one is limited in the variation of the atom-atom interaction 
but can vary the linear coupling between the two wells 
by ramping the potential barrier~\cite{esteve08}. In our 
model, the ground states are determined by the Schr\"odinger
equation in Eq.~(\ref{eq:tr2}). And for the range of values 
of $\Lambda$ to be considered, the parabolic approximation 
is again sufficient, so that for a given $\Lambda$ the ground 
state is, 
\beq
\psi_{\rm GS}(z) = \frac
{1} {[\pi b_{\rm GS}^2]^{1/4}} {\rm e}^{-z^2/(2 b_{\rm GS}^2)}
\eeq
with $b_{\rm GS}^2={4h/\omega}=(2h)/\sqrt{1+\Lambda-h}$.
Retaining terms linear in $h$,
\beqa
\alpha &\simeq&
1+h -{h\over 2\sqrt{1+\Lambda}}\nonumber\\
\xi^2_{N;{\rm GS}}&\simeq&
{1\over \sqrt{1+\Lambda}}
\label{eq:adia}
\eeqa
and thus  
\beq
\xi_{S,{\rm GS}}^2 (\Lambda) =   
{1\over \sqrt{1+\Lambda}} \left[ 1-2h +{h\over \sqrt{1+\Lambda}}\right]
\,.
\label{xigs}
\eeq

\subsection{One-axis twisting ( OAT )}

One-axis twisting was proposed by Kitagawa and 
Ueda~\cite{kita}. Their Hamiltonian is $H_{KU} = \hbar
\chi {\hat J}_z^2$. Compared to Bose-Hubbard, this 
implies that their $J=0$, and $\chi= U$. They worked 
with time, $t_{KU}$, in ``time units'', whereas here 
we express time, $t$, in units of $1/J$. To have more 
compact expressions they introduced $\mu \equiv 2 \chi t_{KU}$: 
in our notation
\begin{equation}
\mu = 2U t_{KU} = 2 U \frac{1}{J} t= \frac{4}{N} \Lambda  t \,.
\label{eq:ku1}
\end{equation}
Since we are here studying squeezings for times of the 
order of the Rabi time, and $N >> 1$, this means that 
in our applications $\mu$ will always be small. 

The initial state considered was $\Psi(\pi/2,0)$ (similar
results are obtained for the $\Psi(\pi/2,\pi)$) so that 
the spin remains aligned along the $x$ axis: 
$\langle{\hat J}_y\rangle = \langle{\hat J}_z\rangle =0$, 
while 
\beq
\langle {\hat J}_x \rangle =  N/2 \cos^{N-1}(\mu/2) \ .
\label{eq:kub2}
\eeq
For small times this simplifies to
\beq
\alpha = 2 h \langle {\hat J}_x \rangle \simeq 1 - 2h \Lambda^2 t^2 \ ,
\label{eq:kub2b}
\eeq
which is the same result found when we expand the 
semiclassical approximation to $\langle{\hat J}_x\rangle$ 
given in Eq.~(\ref{eq:trisa2}). For longer times, in OAT, 
the angle for maximal squeezing was found to be~\cite{kita}
$\beta_{M,{\rm OAT}} = \frac{1}{2} \arctan (B/A) $
with
$A = 1- (\cos \mu)^{N-2}$ and, 
$B = 4 \sin \frac{\mu}{2} \left( \cos \frac{\mu}{2}\right)^{N-2}$. 
The minimum variance in the $(y,z)$ plane is given by, 
\beq
V_-={N\over 4}\left\{\left[ 1+ {1\over4}(N-1)A\right]-{1\over4}(N-1)\sqrt{A^2+B^2}\right\}
\label{eq:kub1}
\eeq
so that 
\beq
\xi_{S,{\rm OAT}}^2= {(4/N) V_-\over \cos^{2(N-1)}(\mu/2)}\,.
\label{eq:kub3}
\eeq
The comparison with the OAT is especially relevant as 
it corresponds to the limit $U\gg J$ of the BH 
Hamiltonian~(\ref{bh01}).

\subsection{Maximal squeezing and scaling properties}

\begin{figure}[t]                         %f01
\begin{center}
\includegraphics[width=8.cm,angle=0]{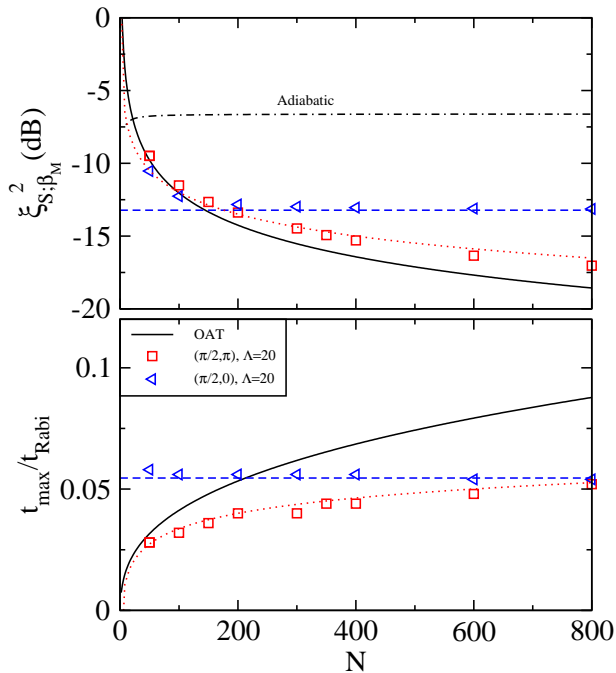}
\caption{(Color online) 
Maximum attainable coherent spin squeezing 
(upper panel) and the time when this maximum value 
is obtained (lower panel) as a function of the number 
of atoms $N$. We compare the methods described in Sections 
$IV$ and $V$ with the adiabatic squeezing, Eq.~(\ref{xigs}) 
(dot-dashed) and the one-axis twisting of Ref.~\cite{kita} 
by means of Eqs.~(\ref{eq:kub3}), and ~(\ref{eq:ku1}) 
(solid lines). The exact Bose-Hubbard calculations 
corresponding to the initial states ($\pi/2,0$) and 
($\pi/2,\pi$) are plotted as triangles and squares, respectively.  
Analytic formulas obtained for the $(\pi/2,\pi)$, Eqs.~(\ref{maxipi})
and Eq.~(\ref{tmax}), are plotted in dotted lines. Analytic 
expressions for the $(\pi/2,0)$ case, the ratio of 
Eq.~(\ref{maxs}) and~(\ref{comaxs}) and the relation above 
Eq.~(\ref{maxs}) which defines the corresponding time, are 
plotted as dashed lines. Note that the plots are made for a 
fixed value $\Lambda=20$.
\label{compmethods}}
\end{center}
\end{figure} 

In Fig.~\ref{compmethods} we compare the maximum attainable 
coherent spin squeezings according to the different methods, 
considering a fixed value of $\Lambda$. First, we note 
that the $N$ scaling of the maximum attainable squeezing 
starting from the $\Psi(\pi/2,0)$ state saturates to 
$\xi^2_{S} \simeq 1/(1+\Lambda)$, with small $1/N$ corrections 
as predicted in Eq.~(\ref{maxs}). This is similar to 
the adiabatic case, which also saturates, albeit to a higher 
value $\xi^2_{S} \simeq 1/\sqrt{1+\Lambda}$.

The large $N$ behavior of the coherent spin squeezing achieved 
from the $\Psi(\pi/2,\pi)$ state is however different. 
The large $N$ scaling of the maximum coherent spin squeezing 
in this case is closer to the one obtained from the one-axis 
twisting method, $\xi_S^2\sim N^{-2/3}$, as seen in 
Fig.~\ref{compmethods} for $\Lambda=20$. In this case, 
the fall-off predicted by Eq.~(\ref{maxipi}) is 
$\xi^2_S \propto (N\Lambda)^{-1/2}$, in good agreement with the
BH results. Two important differences appear however. 
The first one is that these large squeezings are achieved 
at very early times in the evolution of the system, see 
lower panel of Fig.~\ref{compmethods}. Secondly, as shown in 
eq.~(\ref{speed}), the 
parameter $\Lambda$ provides control on the speed of 
coherent spin squeezing in the system.
As seen in Fig.~\ref{compmethods} the time for maximal 
squeezing obtained from the BH calculation is well 
reproduced by Eq.~(\ref{tmax}), showing that the source 
of coherent squeezing in the systems is essentially 
the inflationary parabolic evolution described in 
Section~\ref{sec5}.

Finally let us note that the present results for the time evolution 
of the $(\pi/2,0)$ and 
$(\pi/2,\pi)$ initial states are for moderate $\Lambda= NU/(2J)$ values, i.e. with 
$J\neq 0$. In the $\Lambda\gg 1$ limit the dynamics is the same in both cases,
and, as expected, agrees with that of the OAT. 
Thus, our results are relevant as they quantify the effects 
of the linear coupling $J$ on the maximum coherent spin squeezing achievable
with the considered states.

\section{Summary and conclusions}
\label{sec6}

We have studied the formation of squeezed states 
in the quenched evolution of coherent initial 
states of ultracold atoms trapped in double-well 
potentials. The system is initially prepared in 
either the $(\pi/2,0)$ or $(\pi/2,\pi)$ coherent states, 
which in turn correspond to the ground state 
of the non-interacting system or its highest excited 
state, respectively. 

Simple analytical formulas have been derived which 
correctly describe; a) the dynamics of the system 
for a broad range of repulsive interactions, and, 
b) the formation of squeezed states in the initial 
time evolution. Expressions are given for 
the angle of maximal squeezing and the magnitude 
of the squeezing. The semiclassical model provides 
a mapping relating the dynamical evolution of 
the many-body states considered, to the dynamics 
of a particle evolving on a parabolic potential 
in the Fock-space. Within this picture, the evolution 
of the $\Psi_{(\pi/2,0)}$ state corresponds to that 
of a Gaussian wave packet in the presence 
of a confining parabolic potential, and  
simple periodic formulas describe the time evolution 
of the relevant magnitudes. The evolution of the 
$\Psi_{(\pi/2,\pi)}$ state is mapped, for short times, 
onto the motion of a wave packet in a repulsive 
parabolic potential. In the second case, we have shown 
that the squeezing of the many-body state can be 
much larger than the maximum squeezing obtained in 
the first case, thus providing a promising experimental 
resource for coherent spin squeezing. We have compared 
the maximum attainable squeezing to the adabatic and 
to Kitagawa-Ueda's OAT. We find that the large $N$ 
scaling of the maximum coherent squeezing in the 
$\Psi_{(\pi/2,\pi)}$ case is similar to OAT, but with 
the advantage that the linear coupling $\Lambda$, 
allows to control the speed at which the squeezing 
develops in the system. In the experimentally relevant situation where one is limited by the nonlinearity in the system, this allows to accelerate the generation of squeezing in the system. 

The two initial conditions considered are within reach experimentally in internal bosonic Josephson junctions~\cite{zib10}. We therefore 
expect that the findings reported here will be checked 
against new experiments soon. 

\begin{acknowledgments}
The authors thank J. Tar\'on and M. Lewenstein for useful comments, 
and D. Sprung for a careful reading of the manuscript. This work has been supported by FIS2008-01661 and 
2009-SGR1289. M. M-M. is supported by an FPI 
grant from the MICINN (Spain). B.~J.-D. is supported 
by the Ram\'on y Cajal program. T.Z. acknowledges support from the Landesgraduiertenf\"orderung Baden-W\"urttemberg.

\end{acknowledgments}

%\clearpage

\appendix

\section{Expectation values of 
$\hat{J}_i$ and $\hat{J}_i^2$}
\label{apd}

First note that the action of the spin operators 
${\hat J}_i$ and ${\hat J}_i^2$ on the general state, $|\Psi\rangle$, 
of Eq.~(\ref{eq:ps1}) gives
\beqa
\langle k,N-k| \hat{J}_x \,|\Psi \rangle 
&=&
{1\over 2}  \bigg[ b_k\, c_{k+1} 
+
b_{k-1}\,c_{k-1}\bigg] 
\nonumber \\
\langle k,N-k|\hat{J}_y \,|\Psi \rangle 
&=&
{i\over 2} \bigg[ 
b_{k-1}\, c_{k-1}
-
b_k\, c_{k+1}\bigg] 
\nonumber \\
\langle k,N-k|\hat{J}_z |\Psi \rangle 
&=&
{1\over 2} \left[ 2k -N \right] c_k
\eeqa

and
\beqa
\langle k,N-k| \hat{J}_x^2 \,|\Psi\rangle 
&=&
{1\over4}
\bigg[ 
b_k b_{k+1}\,c_{k+2} 
+
[b_k^2 + b_{k-1}^2]\,c_k 
\nonumber\\
&&+
b_{k-1}b_{k-2}\,c_{k-2}
\bigg]
\\
\nonumber \\
\langle k,N-k| \hat{J}_y^2 \,|\Psi \rangle 
&=&
-{1\over4}
\bigg[ b_k b_{k+1}\,c_{k+2} 
-[b_k^2+b_{k-1}^2]\,c_k \nonumber\\
&&+b_{k-1}b_{k-2}\,c_{k-2}\bigg]
\nonumber \\
\langle k,N-k| \hat{J}_z^2 |\Psi \rangle &=& 
{1\over 4} (2k-N)^2 c_k\nonumber
\eeqa
\beqa
&&\langle k,N-k| \hat{J}_y \hat{J}_z + \hat{J}_z \hat{J}_y  |\Psi \rangle
\\
=&& 
\imath\, \bigg( 
{2k-N-1\over 2} b_k\,c_{k+1}
-
{2k-N+1\over 2} b_{k-1}\,c_{k-1}
\bigg) \nonumber
\eeqa
where $b_k=\sqrt{(k+1)(N-k)}$. We will assume 
that the states, $\Psi$, are such that either 
their $c_k$ vary smoothly (when the initial 
state is $\Psi_{\pi/2,0}$),  or it is their 
$(-)^k c_k$ that vary smoothly (when the initial 
state is $\Psi_{\pi/2,\pi}$.) Also we assume that 
the number of atoms is large, $h = 1/N \ll 1$. This 
allows to introduce a continuous variable , $x$, 
and a continuous function, $\psi(x)$ such that 
$\psi(x=k/N) = \sqrt{N}c_k$ or 
$\sqrt{N} (-)^k c_k$~\cite{Jav99,ST08,martam,ours10-2}.
The factor $\sqrt{N}$ guarantees that $\sum_0^N  |c_k|^2 =1$ 
becomes $\int_0^1 dx |\psi(x)|^2=1$ in the large $N$ limit. 
With these notations and using $b(x)=\sqrt{(x+h)(1-x)}$:

\begin{widetext}
\beqa
c_k^* \langle k,N-k| \hat{J}_x \,|\Psi \rangle 
&=&
\pm {1\over 2} \psi^*(x)  \bigg[ b(x)\,\psi(x+h)
+
b(x-h)\,\psi(x-h)\rangle 
\bigg] \nonumber \\
c_k^* \langle k,N-k|\hat{J}_y \,|\Psi \rangle 
&=&
\pm {\imath \over 2} \psi^*(x) \bigg[ b(x-h)\,\psi(x-h) -b(x)\, \psi(x+h)
\bigg] \nonumber \\
c_k^* \langle k,N-k|\hat{J}_z |\Psi \rangle 
&=&{1\over 2} \psi^*(x) (2x-1)\, \psi(x)
\eeqa
\beqa
c_k^* \langle k,N-k| \hat{J}_x^2 \,|\Psi \rangle 
&=&
{N\over4} \psi^*(x) 
\left[ b(x)b(x+h)\,\psi(x+2h) 
+
\,[b(x)^2+b(x-h)^2]\psi(x) 
+
b(x-h)b(x-2h)\,\psi(x-2h)\right]
\nonumber \\
c_k^* \langle k,N-k|\hat{J}_y^2 |\Psi \rangle 
&=&
-{N\over4} \psi^*(x) 
\left[ b(x)b(x+h)\,\psi(x+2h) 
-
[b(x)^2+b(x-h)^2]\,\psi(x) 
+
b(x-h)b(x-2h)\,\psi(x-2h)
\right] \nonumber \\
c_k^* \langle k,N-k|\hat{J}_z^2 |\Psi\rangle 
&=&
{N\over 4} \psi^*(x) (2x-1)^2 \psi(x)
\eeqa

\beqa
c_k^* \langle k,N-k| \hat{J}_y \hat{J}_z + \hat{J}_z \hat{J}_y  |\Psi \rangle &=& 
\pm {\imath\, N \over 2} \psi^*(x)   \left[ 
(2x-h-1)b(x)\,\psi(x+h)
-
(2x+h-1) b(x-h)\,\psi(x-h)\,,
\right] \nonumber \\
\eeqa
\end{widetext}
where the sign is $+$ ($-$) for states close to $\Psi_{\pi/2,0}$ 
($\Psi_{\pi/2,\pi}$). No approximation has yet been made. Now, we expand 
these expressions in powers of $h=1/N$ up to order $h^2$, 
 introduce the variable $z=2x -1$ and change  
$\psi(x) \to \sqrt{2}\ \psi(z)$ to fulfill:
 $\int_{-1}^1 \ dz |\psi(z)|^2 = 1$. 
In the large $N$ limit, replacing 
the sum over $k$ by an integration over $z$ times $N/2$ one finds:
\begin{widetext}
\beqa
h  \langle \Psi| {\hat J}_x |\Psi \rangle &\simeq& 
 \pm \int_{-1}^1 dz\; \psi^*(z)\,\left[\left(\frac{h^2 \left(-1-z^2\right)}{4 \left(1-z^2\right)^{3/2}}+\frac{h}{2
  \sqrt{1-z^2}}+\frac{\sqrt{1-z^2}}{2}\right) \psi(z) %\nonumber \\&&
-\frac{h^2 z}{\sqrt{1-z^2}}\psi'(z)
+h^2 \sqrt{1-z^2} \psi''(z)\right]\nonumber \\
h  \langle \Psi| {\hat J}_y |\Psi \rangle
&\simeq&  \pm \int_{-1}^{1} dz\; \psi^*(z) \, \bigg[
\left(\frac{\imath h^2 z}{2 \left(1-z^2\right)^{3/2}}-\frac{\imath h z}{2
  \sqrt{1-z^2}}\right) \psi(z)
-2 \left(-\frac{\imath h^2}{2 \sqrt{1-z^2}}
+\frac{\imath h \left(-1+z^2\right)}{2 \sqrt{1-z^2}}\right) \psi'(z) \bigg]
\nonumber \\
h  \langle \Psi| {\hat J}_z |\Psi \rangle
&=& \int_{-1}^1 dz\; \psi^*(z) \,
z \psi(z) \nonumber\\
h^2 \langle \Psi| {\hat J}_x^2 |\Psi \rangle 
&\simeq&
\int_{-1}^1 dz \; \psi^*(z) \,\left[ 
\left(\frac{h}{2}+\frac{1}{4} \left(1-z^2\right)
+\frac{2 h^2 \left(-2+z^2\right)}{8-8 z^2}\right) \psi(z)
-2 h^2 z \psi'(z)
+ h^2 \left(1-z^2\right) \psi''(z)
\right] \nonumber \\
h^2 \langle \Psi| {\hat J}_y^2 |\Psi \rangle 
&\simeq&
\int_{-1}^1 dz \psi^*(z) \,\left[ 
\frac{h^2 \left(-2+z^2\right)}{4 \left(-1+z^2\right)}\psi(z)
+2 h^2 z\psi'(z)
+h^2 \left(-1+z^2\right) \psi''(z)
\right] \nonumber \\
h^2 \langle \Psi| {\hat J}_z^2 |\Psi \rangle 
&=&
\int_{-1}^1 dz\; |\psi(z)|^2 \,{z^2\over 4} \label{eq:trisapendix}
\\
h^2 \langle \Psi| \{{\hat J}_y,{\hat J}_z\} |\Psi \rangle &\simeq& 
 \pm
\int_{-1}^1 dz \,\psi^*(z)\left[ \,
\left(\frac{\imath h}{2 \sqrt{1-z^2}}-\frac{\imath h^2 \left(-1+2 z^2\right)}{2
  \left(1-z^2\right)^{3/2}}\right) \psi(z) %\nonumber \\&&
-2 \left(\frac{\imath h^2 z}{2 \sqrt{1-z^2}}+\frac{1}{2} \imath h z \sqrt{1-z^2}\right)
\psi'(z)
\right] \,.\nonumber
\eeqa
Note that this approximation still fulfills, 
\beq
\langle \Psi| {\hat J}_x^2 +{\hat J}_y^2+{\hat J}_z^2 |\Psi \rangle =
\frac{N}{2}\left(\frac{N}{2}+1\right) \ .
\eeq

\end{widetext}


\begin{thebibliography}{9} 


\bibitem{lewenstein-adv} 
%Ultracold atoms in optical lattices: Mimicking condensed matter and beyond,
M. Lewenstein, A. Sanpera, V. Ahufinger, B. Damski, 
A. Sen (De), and U. Sen, 
Adv. in Phys. {\bf 56}, 243 (2007).%, cond-mat/06006771

\bibitem{bloch-rmp} 
I. Bloch, J. Dalibard, and W. Zwerger, 
%Many-Body physics with ultracold gases. 
Rev. Mod. Phys. {\bf 80}, 885 (2008).

\bibitem{Albiez05}
M. Albiez, R. Gati, J. F\"olling, S. Hunsmann, 
M. Cristiani, and M. K. Oberthaler, 
Phys. Rev. Lett. {\bf 95}, 010402 (2005).

\bibitem{GO07} R. Gati and M. K. Oberthaler, 
%"A bosonic Josephson junction",
 J. Phys. B.: At. Mol. Opt. Phys. {\bf 40}, R61-R89 (2007).

\bibitem{stei07} S. Levy, E. Lahoud, 
I. Shomroni, and J. Steinhauer, 
Nature {\bf 449}, 579 (2007).



\bibitem{esteve08} 
%Squeezing and entanglement in a Bose-Einstein condensate
J. Esteve, C. Gross, A. Weller, S. Giovanazzi, 
and M. K. Oberthaler, Nature {\bf 455}, 1216, (2008).

\bibitem{gross10} C. Gross, T. Zibold, E. Nicklas, 
J. Est\`eve, and M. K. Oberthaler,  
Nature {\bf 464}, 1165 (2010).

% Quantum spin squeezing
\bibitem{spinrev} J. Maa, X. Wanga, C.P. Suna, 
and F. Nori, Phys. Rept. {\bf 509}, 89 (2011).


\bibitem{zib10} 
%Classical Bifurcation at the Transition from Rabi to Josephson Dynamics
T. Zibold, E. Nicklas, C. Gross, 
and M. K. Oberthaler, 
Phys. Rev. Lett. {\bf 105}, 204101 (2010).

\bibitem{Sme97}
A. Smerzi, S. Fantoni, S. Giovanazzi, 
and S. R. Shenoy, 
Phys. Rev. Lett. {\bf 79}, 4950 (1997).

\bibitem{Mil97} G.J. Milburn, J. Corney, 
E. M. Wright, and D. F. Walls, 
%" Quantum
%  dynamics of an atomic Bose Einstein condensate in a double-well potential"
 Phys. Rev. A {\bf 55}, 4318 (1997).

\bibitem{kita} 
M. Kitagawa, and M. Ueda, 
Phys. Rev. A {\bf 47}, 5138 (1993).

\bibitem{korb05} 
J. K. Korbicz, J. I. Cirac, and M. Lewenstein, 
Phys. Rev. Lett. {\bf 95}, 120502 (2005).

\bibitem{wine92} D. J. Wineland, J. J. Bollinger, 
W. M. Itano, and F. L. Moore, and D. J. Heinzen, 
Phys. Rev. A {\bf 46}, R6797 (1992).

\bibitem{wine94} D. J. Wineland, J. J. Bollinger, 
W. M. Itano, and D. J. Heinzen, 
Phys. Rev. A {\bf 50}, 67 (1994).

\bibitem{sinatra11}
A. Sinatra, E. Witkowska, J.-C. Dornstetter, 
Yun Li, and Y. Castin, 
Phys. Rev. Lett. {\bf 107}, 060404 (2011).

%Decoherence as a signature of an excited-state quantum phase transition
\bibitem{rela}
A. Rela\~no, J. M. Arias, J. Dukelsky, 
J. E. Garc\'{\i}a-Ramos, and P. P\'erez-Fern\'andez, 
Phys. Rev. A {\bf 78}, 060102 (2008).

\bibitem{cirac98} 
J. I. Cirac, M. Lewenstein, K. Molmer, and 
P. Zoller, Phys. Rev. A {\bf 57}, 1208 (1998).

\bibitem{jame05} 
M. J\"a\"askel\"ainen, and P. Meystre, 
Phys. Rev. A {\bf 71}, 043603 (2005); 
Phys. Rev. A {\bf 73}, 013602 (2006).

\bibitem{vardi} J. R. Anglin, and, A. Vardi, 
Phys. Rev. A {\bf 64}, 013605 (2001); A. Vardi, 
and J. R. Anglin, Phys. Rev. Lett. {\bf 86}, 568 (2001).

\bibitem{ST08} V. S. Shchesnovich, and 
M. Trippenbach, 
Phys. Rev. A {\bf 78}, 023611, (2008).

\bibitem{ours10} B. Juli\'a-D\'{\i}az, D. Dagnino, 
M. Lewenstein, J. Martorell, and A. Polls, 
Phys. Rev. A {\bf  81}, 023615 (2010).

\bibitem{lipkin} H. J. Lipkin, N. Meshkov, 
and A. J. Glick, Nucl. Phys. {\bf 62}, 188 (1965).

\bibitem{vidal1} P. Ribeiro, J. Vidal, and R. Mosseri, 
Phys. Rev. Lett. {\bf 99}, 050402 (2007).

\bibitem{vidal2} R. Or\'us, S. Dusuel, and J. Vidal, 
Phys. Rev. Lett. {\bf 101}, 025701 (2008).



\bibitem{Leggett01} 
A. J. Leggett, 
%"Bose Einstein condensation in the alkali gases: some
%fundamental concepts"  
Rev. Mod. Phys. {\bf 73}, 307 (2001).

\bibitem{holtaus01} M. Holtaus, and S. Stenholm, 
Eur. Phys. J. B. {\bf 20}, 451 (2001).

\bibitem{Sorensen2001} A. S\o rensen, L.-M. Duan, I. Cirac, 
and P. Zoller, Nature 409, 603 (2001).

\bibitem{ours10-2} B. Juli\'a-D\'{\i}az, 
J. Martorell, and A. Polls, 
Phys. Rev. A {\bf  81}, 063625 (2010). 

\bibitem{dounas07}
%Ultracold Bosons in a Tilted Multilevel Double-Well Potential
D. R. Dounas-Frazer, A. M. Hermundstad, and L. D. Carr, 
Phys. Rev. Lett. {\bf 99}, 200402 (2007).

\bibitem{Jav99} 
J. Javanainen, and M. Yu. Ivanov, Phys. Rev. A {\bf 60}, 2351 (1999).

%Asymptotic approximations to Clebsch-Gordan coefficients from a tight-binding model"
\bibitem{martam}
D.W.L. Sprung, W. van Dijk, J. Martorell, and D.B. Criger, 
Am. J. Phys. {\bf 77}, 552 (2009).


\end{thebibliography}
\end{document}